# Optical properties of Au colloids self-organized into rings via copolymer templates

*S. Lamarre[2], A. Sarrazin[1], J. Proust[1], H. Yockell-Lelièvre[1], J. Plain[1], A. M. Ritcey[2] and T. Maurer[1*]*

*1. Univ Technol Troyes, Lab Nanotechnol & Instrumentat Opt, Inst Charles Delaunay, CNRS UMR 6279, F-10010 Troyes, France*

*2. Univ Laval, Dept Chim, Quebec City, PQ G1V 0A6, Canada*

The investigation of the Localized Surface Plasmon Resonance for plasmonic nanoparticles has opened new perspectives for optical nanosensors. Today, an issue in plasmonics is the development of large scale and low cost devices. We focus here on the Langmuir-Blodgett technique to self-organize gold nanoparticles (~ 7 nm) into rings (~ 60 nm) via polystyrene-b-polymethylmethacrylate templates. In particular, we investigated the optical properties of organized gold nanoparticle rings over large areas and report experimental evidence for plasmon resonances of both individual nanoparticles and collective modes. This paves the way for designing devices with multiple resonances in the visible-Infra-red spectrum and developing optical sensors.

In parallel with progress in characterization techniques, much effort has been made to set up new strategies for smaller metallic nanoparticle synthesis. Two approaches have been developed to synthesize metallic nano-objects: the top-down and the bottom-up ones. The top-down approach consists of taking and etching a massive material down to the desired shape and size. The associated processes, such as Electron Beam Lithography (Haynes et al. 2003) or Focused-Ion-Etching (Tseng 2004), take advantage of high reproducibility in finding immediate applications in the industry. This approach now allows for the fabrication of structures as small as 20 nm in size but it is still challenging to obtain 5-10 nm structures with well-defined edges via these techniques (Pelton et al. 2008). In order to go below this threshold, synthetic routes, like colloidal synthesis (Sun and Xia 2002), based on the *bottom-up* approach have been developed for about twenty years. The great advantage of this approach is that it provides significant amounts of nano-objects with dimensions smaller than 20 nm or even 10 nm (Soumare et al 2009). Nevertheless, for the moment, large-scale applications are still limited with this approach due to the difficulty of organizing such metallic nano-objects over large areas. Therefore, a big challenge today is to develop routes for organizing colloids onto substrates, especially for non magnetic nanoparticles (Maurer et al. 2007). Among the most appealing strategies for industrial applications, is that based on the use of block copolymers since they exhibit an exceptional ability to self-organize into microphase-separated domains with controlled shapes and sizes (Fahmi et al. 2009; Kim et al. 2004). Indeed for lithography, block copolymers can provide not only long-range ordered templates but also small feature size and a large variety of patterns, ranging from spheres to lamellar shapes (Tseng and Darling 2010, Krausch and Magerle 2002). The Reactive Ion Etching (RIE) rate of polymethylmethacrylate (PMMA) is sufficiently lower for $CF_4$ or $O_2$ gas compared to that of polystyrene (PS) to allow the selective removal of PMMA in PS-*b*-PMMA block copolymer templates (Asakawa and Hiraoka 2002, Asakawa and Fujimoto 2005). It has also been shown for PS-*b*-PMMA block copolymer templates that PMMA can be selectively degraded after UV-exposure and then removed via immersion in acetic acid solution (Kang et al 2009). All of these processes are bottom-up strategies to prepare lithography masks with long-range ordered pores. Depending on the depth of the holes, metal nanoparticles can be obtained either after metal evaporation and resin removal (Kang et al. 2009), after surface substrate functionalization and colloid deposition (Gowd et al. 2009; Choi et al. 2009) or after electrophoretic deposition of the colloidal particles (Zhang et al. 2005). Fully bottom-up strategies have also been developed from block copolymers to synthesize nanocomposite thin films. In particular, the efforts of many researchers have been focused onto the development of single-step synthesis and organization of inorganic nanoparticles in block copolymer templates. Indeed, inorganic precursors can be selectively incorporated into one block of amphiphilic block copolymers like polystyrene-*b*-poly(2-vinylpyridine) [Spatz et al. 1996, Mössmer et al 2000]. For exemple, poly(ethylene oxide)-*b*-poly(propylene oxide) can efficiently reduce and stabilize inorganic precursors like trihydrate tetrachloroaureate ($HAuCl_4 \cdot 3H_2O$), leading to the single step synthesis of gold nanoparticles (Sakai and Alexandridis 2004, Alexandridis 2011).

The scope of this paper is the investigation of the optical properties for composite films with Au colloids self-organized into rings via copolymer templates. Nanorings have attracted much attention for the past ten years (Aizpurua et al. 2003, Nordlander 2009, Lerond et al. 2011) with the development of colloidal lithography. Such structures are characterized by collective plasmon modes determined by ring geometry rather than the geometry of individual nanoparticles (Nordlander 2009). Therefore, they provide appealing features such as resonances in the near Infra-Red (IR) spectrum (Aizpurua et al. 2003, Lerond et al. 2011). The originality of this paper is to focus on rings which are much smaller than those provided by colloidal lithography and composed of only a relatively small number (~ 10) of individual nanoparticles. We first describe self-organization of previously functionalized gold nanoparticles into domains of PS-*b*-PMMA via the Langmuir-Blodgett process (Lamarre et al. 2013). We then investigate optical properties of such hybrid functional materials and provide spectral evidence for plasmonic interaction between the nanoparticles. Experimental results are discussed and compared to numerical simulations. Perspectives and possible applications for such hybrid materials will be discussed in conclusion.

The synthesis of the nanoparticles investigated in this study is based on the phase transfer reduction protocol introduced by Brust and coworkers (Brust et al. 1994). The exact procedure was previously detailed by H. Yockell-Lelièvre et al. (Yockell-Lelièvre et al. 2007). Tetraoctylammonium bromide (TOABr) was used as a phase transfer agent and, unlike the original Brust synthesis, thiolated octane chains were added in a second step via a ligand exchange reaction. Octanethiol was purchased to Aldrich. Direct chemisorption of the thiolated octane chains on the surface of the phase transfer reduced NPs was carried out by the dropwise addition of 10 mL of the organosol to an excess of thiolated chains in chloroform (1 g/10 mL) under constant stirring. The system was stirred at room temperature for 24 h to ensure surface saturation with the thiol. The samples were then purified by repeated fractional reprecipitation-centrifugation cycles in a chloroform/methanol solution. Infrared analysis of the purified samples revealed no residual phase transfer agent. In the dry state, all samples were stable, and no changes in properties were noted with time (Yockell-Lelièvre et al. 2007). TEM images of the synthesized gold nanoparticles were obtained with a Jeol JEM-1230 Transmission Electron Microscope and indicated an average diameter of 7nm± 1nm.

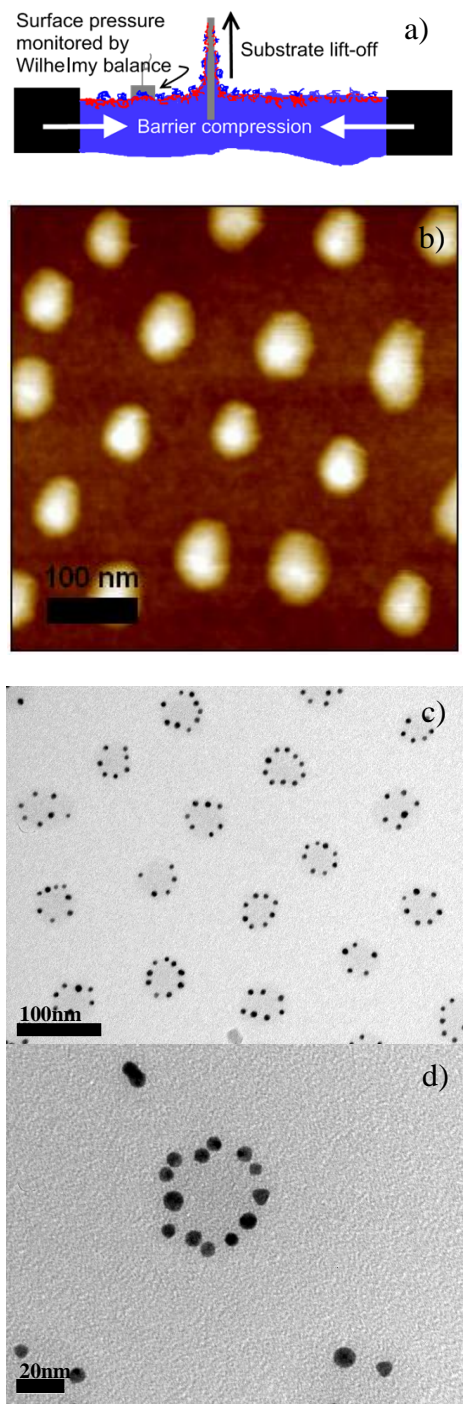

**Fig.1** a) Schematic representation of the Au nanoparticle-copolymer film spread onto water in the Langmuir-Blodgett set-up. The PS block forms circular aggregates whereas the PMMA block (red) lies on the water surface due to its higher affinity. b) Atomic force microscopy image of PS-*b*-PMMA monolayer deposited on glass substrate by LB transfer. c) TEM image of Au nanoparticles organized in PS domains. d) TEM image illustrating the tendency of the Au nanoparticles to form rings at the interface between the PS domains and the PMMA matrix.

The Au nanoparticles were dispersed in a copolymer solution of PS-*b*-PMMA ($M_n$=104 000 g/mol, 50 000 and 54 000 for PS and PMMA blocks, respectively) in chloroform which was then spread at the air-water interface in a Langmuir-Blodgett set-up (see **Fig. 1a**). The process has been detailed by Lamarre and coworkers (Lamarre et al. 2013). The process is here based on the higher affinity of water for PMMA than for PS. Therefore, spreading the composite copolymer solution leads to the formation of a PMMA thin film on the water surface decorated with PS islands (see **Fig. 1b**). Depending on both the size of the nanoparticles and the surface ligand (Lamarre et al. 2013), the nanoparticles can either aggregate (10 nm larger or coated with PS nanoparticles), be evenly dispersed within the PS domains (2 nm coated with alcanethiol chains nanoparticles) or be at the air/PS/PMMA triple interface (6 nm nanoparticles coated with $C_8$-SH chains). Once deposited on solid substrate these features are stable at least for months since we do not observe aggregation of the nanoparticles. Our investigations focused on the last case for which Au nanoparticles self-organize into rings (see **Figs. 1c and d**). It is possible to tune ring diameter by selecting PS-*b*-PMMA samples with different molar masses or by swelling PS domains by adding homopolymer PS to the composite, although these possibilities are not pursued in this article.

The extinction spectra of the systems have been measured with a transmission optical microscope coupled to a micro-spectrometer by a multimode optical fibre as schematically described in **Fig. 2a**. A ×100 objective lens (NA = 0.9) is coupled to a tube lens with a focal length of 150mm instead of 180mm which conducts to a x83 magnification. The spatial filtering is granted by a 62.5µm large core optical fiber which is confocal to the sample. This optical set-up allows for a detection area of ≈0.45 µm² which corresponds to about 35 domains filled with Au nanoparticle rings. The collected zone can be observed and spatially adjusted by injecting a laser illumination from the other extremity of the optical fiber which gives rise to its image on the sample. **Fig. 2b** shows the extinction spectrum under normal incidence and clearly exhibits two peaks at $\lambda_1$=525nm and $\lambda_2$=615nm. A third peak may be suspected at $\lambda_3$~750nm. These experimental extinction spectra deserve attention since they exhibit plasmon modes at wavelengths much larger than expected for such small gold nanoparticles.

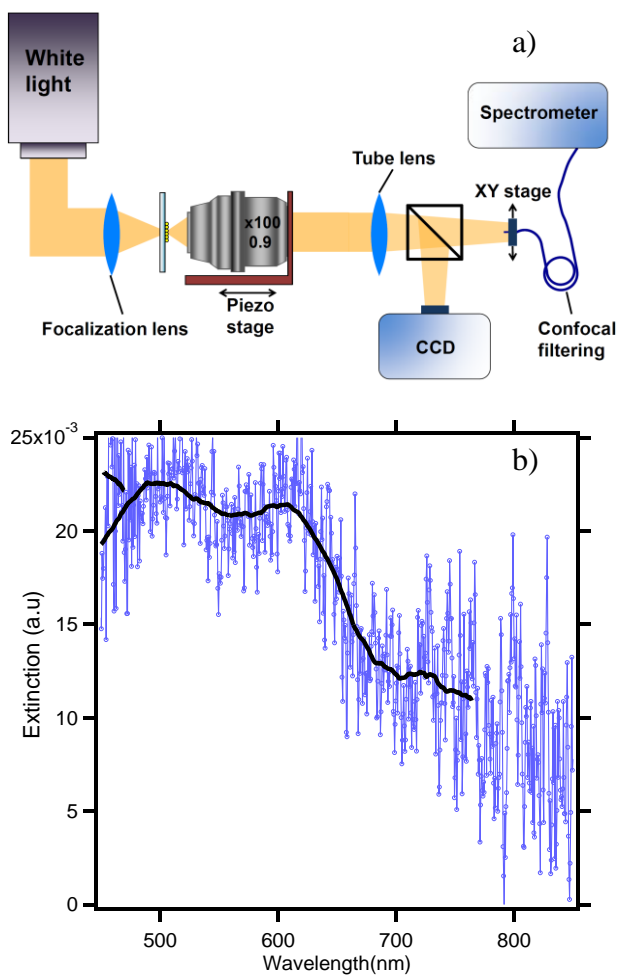

**Fig.2** a) Set up for optical extinction measurements b) Extinction spectrum (purple line and markers) of gold nanoparticles organized into rings. The probed surface is around 0.65µm² which corresponds to about 50 PS domains.

The peak at $\lambda_1$=525nm is indeed attributed to the Localized Surface Plasmon Resonance of single 6 nm gold nanoparticles as predicted by the Mie theory for PS as surrounding medium (Bohren and Huffman 1983). However, the other peaks at $\lambda_2$ and $\lambda_3$ may come from collective resonances due to the ring structures as previously evidenced in studies on Au rings synthesized by colloidal lithography (Aizpurua et al. 2003, Lerond et al. 2011). Numerical simulations performed with the Discrete Dipole Approximation (DDA) modelled extinction spectra of twelve 6 nm gold nanoparticles organized into rings in polystyrene matrix (see **Fig. 3a**). The results clearly show that for such small nanoparticles, additional peaks for incident wavelengths larger than 600 nm only appear when nanoparticles are spatially very close together (ring diameters inferior to 26 nm). The extinction spectra obtained experimentally and showing several plasmon modes are thus a signature of rings with close packed nanoparticles (see **Fig. 1c**) and rings with well separated nanoparticles are characterized by the mode at $\lambda_1$. When the nanoparticles are almost in contact (ring diameters of 23.2 nm and 24 nm, see **Fig. 3a**, a mode appears whose wavelength is even larger than 700 nm. This implies that by tuning the geometry of the rings, it is possible to tune the extinction wavelength of these plasmonic structures. To go further, numerical simulations via the Finite Difference Time Domain (FDTD) method were performed to obtain the extinction spectrum of gold nanorings with a 50 nm internal diameter and a 7 nm thickness (see **Fig. 3b**). This simulation evidences three main modes in the visible range, at 525 nm, 650 nm and 750 nm which almost corresponds to the experimental results. It implies that when Au nanoparticles tends to form rings, their plasmon modes become determined by ring geometry rather than the geometry of individual nanoparticles as already evidenced by Nordlander 2009.

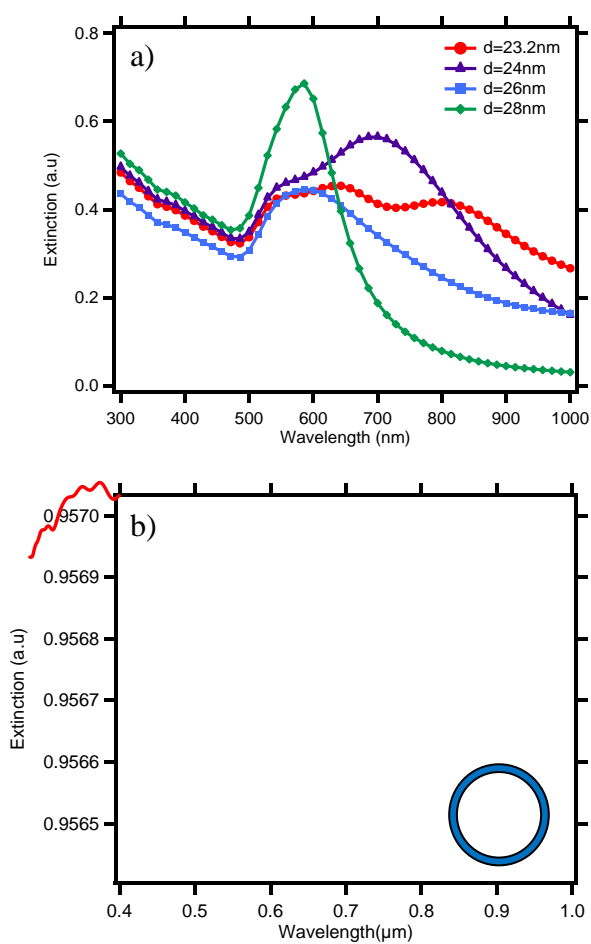

**Fig.3** a) Calculated extinction spectra of twelve 6 nm Au spheres organized into rings via the Discrete Dipole Approximation. The extinction spectra were modeled for different values of the ring diameter: 23.2nm (red circles), 24nm (purple triangles), 26nm (blue squares) and 28nm (green rhombs). The incident light is supposed normal to the ring. b) Calculated extinction spectrum of an Au nanoring via the Finite Difference Time Domain (FDTD) method. The internal diameter of the nanoring is 50nm and its thickness is 7nm. The extinction spectrum exhibits peaks at the following wavelengths in the visible range: 525nm, 650nm and 750nm.

In conclusion, the Langmuir-Blodgett technique is a suitable technique to provide large scale organization of small gold colloids organized into rings in the diameter range 25-60 nm. These plasmonic structures are of primary interest since they exhibit features of both individual gold nanoparticles and plasmonic rings. Therefore, it is possible to tune the optical resonances from visible to near IR spectrum but also to design devices with multiple resonances in the visible-IR range. This could have significant applications in the field of Localized Surface Plasmon Resonance (LSPR) and Surface Enhanced Raman Spectroscopy (SERS) sensors in order to detect molecules. We believe that attention should be focused on the use of such structures for biosensing in future studies.

**Acknowledgments**

Financial support of NanoMat (www.nanomat.eu) by the "Ministère de l'enseignement supérieur et de la recherche," the "Conseil régional Champagne-Ardenne," the "Fonds Européen de Développement Régional (FEDER) fund," and the "Conseil général de l'Aube" is acknowledged. The authors thank the DRRT (Délégation Régionale à la Recherche et à la Technologie) of Champagne-Ardenne, le Fonds Québécois de la recherche sur la nature et les technologies (FQRNT) and the National Sciences and Engineering Research Council of Canada (NSERC) for financial support. T. M. also thanks the CNRS via the chair « optical nanosensors » and the Labex ACTION project (contract ANR-11-LABX-01-01) for financial support.